\documentclass[doublecol]{epl2} 
\usepackage{graphicx} 
\usepackage{epsfig} 
\usepackage{dcolumn}
\usepackage{bm}
\usepackage{color}

\newcommand{\be}{\begin{eqnarray}}
\newcommand{\ee}{\end{eqnarray}}

\def\beq{\begin{equation}}
\def\eeq{\end{equation}}
\def\beqa{\begin{eqnarray}}
\def\eeqa{\end{eqnarray}}
\def\ban{\begin{eqnarray*}}
\def\ean{\end{eqnarray*}}
\def\bi{\begin{itemize}}
\def\ei{\end{itemize}}

\title{Estimation of the effect of hyperonic three-body forces on the maximum mass of neutron stars}

\author{I. Vida\~na\inst{1}\footnote{Corresponding author: ividana@fis.uc.pt} \and D. Logoteta\inst{1} \and C. Provid\^encia\inst{1} \and A. Polls\inst{2} \and I. Bombaci\inst{3}}

\institute{                    
  \inst{1} Centro de F\'{i}sica Computacional, Department of Physics, University of Coimbra, PT-3004-516, Coimbra, Portugal\\
  \inst{2} Departament d'Estructura i Constituents de la Mat\`eria and Institut de Ci\`{e}ncies del Cosmos, Universitat de Barcelona, Avda. Diagonal 647, E-08028 Barcelona, Spain \\
  \inst{3} Dipartimento di Fisica ``E. Fermi'', Universit\`a di Pisa, and INFN, Sezione di Pisa, Largo B. Pontecorvo 3, I-56127 Pisa, Italy
  }


\abstract{
A model based on a microscopic Brueckner--Hartree--Fock approach of hyperonic matter supplemented with additional simple
phenomenological density-dependent contact terms is employed to estimate the effect of hyperonic three-body forces on the
maximum mass of neutron stars. 
Our results show that although hyperonic three-body forces can reconcile
the maximum mass of hyperonic stars with the current limit of $1.4-1.5 M_\odot$, they are unable
to provide the repulsion needed to make the maximum mass compatible with the observation of massive neutron stars,
such as the recent measurements of the unusually high masses of the millisecond pulsars PSR J1614-2230 ($1.97 \pm 0.04 M_\odot$) 
and PSR J1903+0327 ($1.667 \pm 0.021 M_\odot$).}

\begin{document}

\maketitle

Neutron stars are the remnants of the gravitational collapse of massive stars during a Type II, Ib or Ic
supernova explosion. Their masses and radii are typically of the order of $1-2 M_\odot$ ($M_\odot \simeq 2 \times 10^{33}$ g
is the mass of the Sun) and $10-12$ km, respectively. With central densities in the range of $4-8$ times the normal 
nuclear matter saturation density, $\epsilon_0 \sim 2.7 \times 10^{14}$ g/cm$^3$ ($\rho_0 \sim 0.16$ fm$^{-3}$), neutron stars are most likely among
the densest objects in the universe \cite{shapiro83}. Nowadays, it is still an open question which is the true nature of neutron stars.
Traditionally the core of neutron stars has been modeled as an uniform fluid of neutron rich nuclear matter in equilibrium 
with respect to the weak interaction ($\beta$-stable nuclear matter). Nevertheless, due to the large value of
the density, new hadronic degrees of freedom are expected to appear in addition to nucleons. Hyperons, baryons with
a strangeness content, are an example of these degrees of freedom. Since the pioneer work of Ambartsumyan and Saakyan 
\cite{ambar60} the presence of hyperons in neutron stars has been studied by many authors using either phenomenological 
\cite{glendenning85,weber89,schaffner96,balberg97} or microscopic \cite{baldo00,vidana00,hans06,dapo10} approaches. Hyperons 
may appear in the inner core of neutron stars at densities of about $2-3 \rho_0$. At such densities, 
the nucleon chemical potential is large enough to make the conversion of nucleons into hyperons energetically favorable. 
This conversion relieves the Fermi pressure exerted by the baryons, and makes the equation of state (EoS) softer. As a 
consequence, the maximum mass of the star is substantially reduced to values that, in microscopic calculations 
\cite{baldo00,vidana00,hans06,dapo10}, can be even below the ``canonical'' one of $1.4-1.5 M_\odot$ inferred from precise neutron star mass 
determinations \cite{thorsett99}. It has been suggested, see {\it e.g.} Ref. \cite{takatsuka}, that a possible solution to this problem is the inclusion 
of three-body forces (TBF) involving one or more hyperons ({\it i.e.,} nucleon-nucleon-hyperon (NNY), nucleon-hyperon-hyperon (NYY), and 
hyperon-hyperon-hyperon (YYY)). These forces could eventually provide the additional repulsion needed to make the EoS stiffer and, therefore, the maximum 
mass compatible with the current observational limits. In this letter, we want to establish numerical lower and upper limits to the effect of these 
forces on the maximum mass of neutron stars that can guide more sophisticated calculations in the future. To such end, we use a model 
based on a microscopic Brueckner--Hartree--Fock (BHF) approach of hyperonic matter \cite{baldo00,vidana00,hans06,bhf} supplemented with additional 
simple phenomenological density-dependent contact terms that account for the effect of nucleonic and hyperonic TBF.

Our calculation starts by constructing all the baryon-baryon $G$ matrices, which describe the interaction between 
two baryons in the presence of a surrounding medium. The $G$ matrices are obtained by solving the coupled-channel Bethe--Goldstone equation, written   
schematically as
\begin{eqnarray}
G(\omega)_{B_1B_2,B_3B_4}=V_{B_1B_2,B_3B_4} +\sum_{B_iB_j}V_{B_1B_2,B_iB_j} \nonumber \\
\times \frac{Q_{B_iB_j}}{\omega-E_{B_i}-E_{B_j}+i\eta}G(\omega)_{B_iB_j,B_3B_4} \ ,
\label{eq:bge}
\end{eqnarray}
where the first (last) two subindices indicate the initial (final) two-baryon states compatible with a given value
$S$ of the strangeness, namely nucleon-nucleon (NN) for $S=0$ and hyperon-nucleon (YN) for $S=-1$; $V_{B_1B_2,B_3B_4}$ is the bare baryon-baryon
interaction (NN$\rightarrow$NN, $\Lambda$N$\rightarrow$$\Lambda$N, $\Lambda$N$\rightarrow$$\Sigma$N or $\Sigma$N$\rightarrow$$\Sigma$N ); 
$Q_{B_iB_j}$ is the Pauli operator, that prevents the intermediate
baryons $B_i$ and $B_j$ from being scattered to states below their respective Fermi momenta; and the starting energy
$\omega$ corresponds to the sum of the nonrelativistic single-particle energies of the interacting baryons. 

The single-particle energy of a baryon $B_i$ is given by
\begin{equation}
E_{B_i}(\vec{k})=M_{B_i}+\frac{\hbar^2k^2}{2M_{B_i}}+\mbox{Re}[U_{B_i}(\vec{k})]\ .
\label{eq:spe}
\end{equation}
Here $M_{B_i}$ denotes the rest mass of the baryon, and the single-particle potential $U_{B_i}$ represents 
the average field ``felt'' by the baryon owing to its interaction with the other baryons of the medium. In the BHF approximation,
$U_{B_i}$ is calculated through the ``on-shell energy'' $G$-matrix, and is given by
\begin{eqnarray}
U_{B_i}(\vec{k})= \sum_{B_j}\sum_{\vec{k'}}n_{B_j}(|\vec{k'}|) \hspace{1.0cm} \nonumber \\
\times \langle \vec{k}\vec{k'}|G(E_{B_i}(\vec{k})+E_{B_j}(\vec{k'}))_{B_iB_j,B_iB_j} |\vec{k}\vec{k'} \rangle_{\cal A} \ ,
\label{eq:spp}
\end{eqnarray}
where $n_{B_j}(|\vec{k}|)$ is the occupation number of the species $B_j$, and the index ${\cal A}$ indicates that the matrix elements are properly 
antisymmetrized when baryons $B_i$ and $B_j$ belong to the same isomultiplet. We note here that the so-called continuous prescription has been adopted for 
the single-particle potentials when solving the Bethe--Goldstone equation, since, as shown by the authors of Ref.\ \cite{song98}, the contribution to the 
energy per particle from three-hole line diagrams is minimized in this prescription.  
All the calculations carried out in this letter have been performed with the realistic Argonne V18 \cite{wiringa95} NN force and the Nijmegen 
soft-core NSC89 \cite{maessen89} YN interaction. 
We recall that the NSC89 potential does not contain hyperon-hyperon (YY) components.

Once a self-consistent solution of Eqs.\ (\ref{eq:bge})--(\ref{eq:spp}) is obtained, the
energy density can be calculated in the BHF approximation as
\begin{equation}
\epsilon_{BHF}=\frac{1}{{\cal V}}\sum_{B_i}\sum_{\vec{k}}n_{B_i}(|\vec{k}|)\left[
M_{B_i}+\frac{\hbar^2k^2}{2M_{B_i}}+\frac{1}{2}U_{B_i}(\vec{k})\right] \ .
\label{eq:ev}
\end{equation}  

TBF can be introduced in the BHF approach by adding effective density-dependent two-body forces to the baryon-baryon 
interactions $V$ when solving the Bethe--Goldstone equation for the $G$-matrices. These effective forces are obtained by averaging 
genuine baryon-baryon-baryon TBF, $W_3({\vec r}_i,{\vec r}_j,{\vec r}_k)$, over the third baryon coordinates \cite{loiseau71},
\begin{equation}
V^{eff}_{B_iB_j}({\vec r}_{ij})=\int W_3({\vec r}_i,{\vec r}_j,{\vec r}_k)n({\vec r}_i,{\vec r}_j,{\vec r}_k)d{\vec r}_k \ ,  
\label{eq:tbf1}
\end{equation}
where ${\vec r}_{ij}$ is the relative coordinate of baryons $B_i$ and $B_j$, and $n({\vec r}_i,{\vec r}_j,{\vec r}_k)$ is an appropriate 
three-body correlation function. This procedure has been the usual one to introduce nucleonic 
TBF in BHF and other non-relativistic many-body approaches of nuclear matter (see {\it e.g.,} Ref.\ \cite{zhou04}). 
Nevertheless, since there is an 
almost complete lack of experimental data, and few theoretical studies \cite{chalk63,nyman67,gal67,bhaduri67,loiseau69,bodmer84,yamamoto87,kimura92,usmani99}, 
on the genuine hyperonic TBF, instead of using the average procedure described, in this letter we adopt an alternative strategy. First, we 
construct, as described above, the hyperonic matter EoS within the BHF approach using only two-body NN and YN forces, and then, we add to it simple 
phenomenological density-dependent contact terms that account for the effect of both nucleonic and hyperonic TBF.  The reader could think that in doing that 
the predictive power of our model is partially lost. However, we would like to stress once more that the aim of this work is to perform a simple estimation of 
the effect of hyperonic TBF on the maximum mass of neutron stars that can guide and motivate more realistic and sophisticated studies 
of such forces and their effects. We want also to point out that by using a simple parametrization of the hyperonic TBF we can easily 
explore whether or not such forces can make the EoS stiff enough so that the maximum mass of hyperonic stars can be reconciled with the 
observational limits. Following Balberg and 
Gal (see Eqs.\ (2) and (3) of Ref.\ \cite{balberg97}), we assume these density-dependent contact terms to have a Skyrme-like form
\begin{equation}  
\epsilon_{xy}(\rho_x,\rho_y)=a_{xy}\rho_x\rho_y+b_{xy}\rho_x\rho_y\left(\frac{\rho_x^{\gamma_{xy}}+\rho_y^{\gamma_{xy}}}{\rho_x+\rho_y} \right) \ ,
\label{eq:ct}
\end{equation}
where $x$ and $y$ denote any baryon species. Note that, to be precise only the second term actually accounts for the effect of three- and multi-body forces 
which are responsible for the repulsion that dominates at high densities. The first one is in fact a two-body term that modifies slightly the BHF part of 
the energy density by yielding additional attraction at low densities. It can be considered as a rearrangement of the two-body contribution due to the inclusion 
of TBF. It mimics the modification that the effective density-dependent two-body force, $V^{eff}_{B_iB_j}$, would introduce on the $G$-matrices. Then, we have 
the following contact terms contributing to the energy density
\begin{eqnarray}  
\epsilon_{CT}  &=& a_{NN}\rho_N^2+b_{NN}\rho_N^{\gamma_{NN}+1} \nonumber \\
&+&a_{\Lambda N}\rho_\Lambda\rho_N+b_{\Lambda N}\rho_\Lambda\rho_N\left(\frac{\rho_\Lambda^{\gamma_{\Lambda N}}+\rho_N^{\gamma_{\Lambda N}}}{\rho_\Lambda+\rho_N} \right) \nonumber \\
&+&a_{\Sigma N}\rho_\Sigma\rho_N+b_{\Sigma N}\rho_\Sigma\rho_N\left(\frac{\rho_\Sigma^{\gamma_{\Sigma N}}+\rho_N^{\gamma_{\Sigma N}}}{\rho_\Sigma+\rho_N} \right) \ , \nonumber \\
\label{eq:tbf2}
\end{eqnarray}
where $\rho_N=\rho_n+\rho_p$, $\rho_\Sigma=\rho_{\Sigma^-}+\rho_{\Sigma^0}+\rho_{\Sigma^+}$, and we assume charge independence.  
We note that the first two terms mime the contribution of the effective two-body interation, $V^{eff}_{NN}$, obtained when the
forces NNN and NNY are averaged over the coordinates of one of the nucleons and the hyperon, respectively (see Eq.\ (\ref{eq:tbf1})). 
Similarly, the remaining terms mimic the contributions of the effective $V^{eff}_{YN}$ forces built from 
the average of the NNY and NYY ones over the coordinates of the appropriate nucleons and hyperons. 
Note that the NYY forces can also contribute, 
in addition to the YYY ones, to the effective density-dependent YY forces if the average is taken over the nucleon coordinates.
For consistency with the use of the NSC89 YN potential, which as it is said, does not describe any YY interaction, none of the terms of Eq.\ (\ref{eq:tbf2}) 
mimic neither this other contribution of NYY, nor the ones of YYY. 
The complete baryonic energy density is then  simply obtained by adding 
$\epsilon_{CT}$ to the BHF energy density of Eq.\ (\ref{eq:ev}). 

The parameters $a_{NN}$, $b_{NN}$ and $\gamma_{NN}$ are fitted to reproduce the density, $\rho_0$, energy per particle, $E/A$, and incompressibility coefficient, $K_\infty$, 
of symmetric nuclear matter at saturation. We choose the saturation point to be at $\rho_0=0.16$ fm$^{-3}$ and $E/A=-16$ MeV. 
Due to the uncertainty still existing on 
the precise value of $K_\infty$, we consider four values of $\gamma_{NN}$ between $2$ and $3.5$, giving $K_\infty$ in the range $211-285$ MeV, compatible with 
the value of $K_\infty=240 \pm 40$ MeV supported nowadays by experimental data \cite{youngblood99}. The values of $a_{NN}$, $b_{NN}$, $\gamma_{NN}$ and $K_\infty$ 
are given in Table \ref{tab:tab1}. We note that due to the charge independence assumed in Eq.\ (\ref{eq:tbf2}) neither the symmetry energy, $E_{sym}(\rho)$, nor its 
derivative parameter $L=3\rho_0\partial E_{sym}(\rho)/\partial \rho |_{\rho_0}$ will be affected by the inclusion of the contact terms. Their values at $\rho_0=0.16$ fm$^{-3}$
are $28.3$ and $46.6$ MeV, respectively.

\begin{center}
\begin{table}[t]
\begin{tabular}{cccc}
\hline
\hline
$\gamma_{NN}$ & $a_{NN}$ & $b_{NN}$ & $K_\infty$ \\
              & [MeV fm$^3]$ & [MeV fm$^{3\gamma_{NN}}$] & [MeV] \\
\hline
$2$   &  $-33.44$ & $213.02$ &  $211$ \\ 
$2.5$ &  $-22.08$ & $355.03$ &  $236$ \\ 
$3$   &  $-16.40$ & $665.68$ &  $260$ \\ 
$3.5$ &  $-12.99$ & $1331.36$ &  $285$ \\ 
\hline
\hline
\end{tabular}
\caption{Values of the parameters $a_{NN}$, $b_{NN}$, $\gamma_{NN}$, and the incompressibility coefficient $K_\infty$.}
\label{tab:tab1}
\end{table}
\end{center}
In order to reduce the number of parameters, in this exploratory work, we assume for simplicity that TBF involving $\Lambda$ 
and $\Sigma$ hyperons are the same. Therefore, we take $a_{\Lambda N}=a_{\Sigma N}, b_{\Lambda N}=b_{\Sigma N}$ and 
$\gamma_{\Lambda N}=\gamma_{\Sigma N}$. The parameters $a_{YN}$ and $b_{Y N}$ ($Y=\Lambda, \Sigma$) are considered free.
We take them to be a fraction $x$ of the parameters $a_{NN}$ and $b_{NN}$ ($a_{YN}=x\,a_{NN}$ and $b_{YN}=x\,b_{NN}$ 
with $x=0, 1/3, 2/3$ and $1$) in order to explore different strengths of the hyperonic TBF. 
There is no physical reason for which these parameters should scale 
with the same factor $x$ with respect to their nucleonic counterparts. However, we have checked 
that allowing different scaling factors for $a_{YN}$ and $b_{YN}$, we always obtain maximum masses 
within the range of values shown in Table \ref{tab:tab2}. As an example, we note that for $\gamma_{NN}$=3.5 
we obtain $M_{max} = 1.61 M_\odot$ for $a_{YN}=a_{NN}/3$ and $b_{YN}=b_{NN}$, which is very similar to the 
value of $1.60 M_\odot$ shown in Table \ref{tab:tab2} for $x=1$. In view of this, and for simplicity, we keep 
the same scaling factor $x$ for both $a_{YN}$ and $b_{YN}$. 
Finally, the value of $-28$ MeV for the binding 
energy of a single $\Lambda$ in nuclear matter at saturation, extracted from the extrapolation of hypernuclear data 
\cite{millener88}, is then used to determine the parameter $\gamma_{Y N}$ through the relation
\begin{equation}
\left(\frac{B}{A}\right)_\Lambda=-28 \,\, \mbox{MeV} =U_{\Lambda}(k=0)+a_{YN}\rho_0+b_{YN}\rho_0^{\gamma_{YN}} \ , 
\label{eq:byn}
\end{equation}
where $U_{\Lambda}(k=0)$ ($-30.8$ MeV in our calculation) is the BHF single-particle potential of the $\Lambda$ (see Eq.\ (\ref{eq:spp}))
at zero momentum, and the last two terms account for the contact term contribution.

Once the total energy density $\epsilon=\epsilon_{BHF}+\epsilon_{CT}+\epsilon_L$ ($\epsilon_L$ being the contribution of 
noninteracting leptons), is known, the composition and the EoS of neutron star matter can be obtained from the requirement of 
equilibrium under weak interaction processes, $\mu_i=b_i\mu_n-q_i\mu_e$ ($b_i$ and $q_i$ denoting the baryon number and charge 
of species $i$) and  electric charge neutrality, $\sum_i q_i\rho_i=0$. The chemical potentials of the various species and the 
pressure are computed from the usual thermodynamical relations, $\mu_i=\partial \epsilon/\partial \rho_i$ and 
$P=\rho^2\partial (\epsilon/\rho)/\partial \rho$. Finally, knowing the EoS, the equilibrium configurations of static neutron stars 
are obtained by solving the well-known Tolman--Oppenheimer--Volkoff (TOV) equations \cite{shapiro83}.
\begin{center}
\begin{table}[t]
\begin{tabular}{cccccc}
\hline
\hline
$\gamma_{NN}$ & $x$  & $\gamma_{YN}$ & $M_{max}$ & $\rho_c$ & $v_s$  \\
\hline
      &  0 & - & $1.27 \, (2.22)$ & $1.35 \,  (1.07)$ & $0.46 \, (1.03)$ \\ 
      &  $1/3$ & $1.49$ &  $1.33$ & $1.33$ & $0.48$ \\ 
$2$   &  $2/3$ & $1.69$ &  $1.38$ & $1.29$ & $0.52$ \\ 
      &  $1$   & $1.77$ &  $1.41$ & $1.24$ & $0.54$ \\ 
\hline
      &  0 & - & $1.29 \, (2.46)$ & $1.19 \, (0.92) $ & $0.43 \, (1.17)$  \\ 
      &  $1/3$ & $1.84$ &  $1.38$ & $1.16$ & $0.49$ \\ 
$2.5$ &  $2/3$ & $2.08$ &  $1.44$ & $1.12$ & $0.54$ \\ 
      &  $1$   & $2.19$ &  $1.48$ & $1.09$ & $0.56$ \\ 
\hline
      &  0 & - & $1.34 \, (2.72)$ & $0.98 \, (0.79)$ & $0.40 \, (1.34)$ \\ 
      &  $1/3$ & $2.23$ &  $1.45$ & $0.97$ & $0.50$ \\ 
$3$   &  $2/3$ & $2.49$ &  $1.50$ & $0.94$ & $0.55$ \\ 
      &  $1$   & $2.62$ &  $1.54$ & $0.90$ & $0.58$ \\ 
\hline
      &  0 & - & $1.38 \, (2.97)$ & $0.87 \,(0.69)$ & $0.38 \, (1.47)$  \\ 
      &  $1/3$ & $2.63$ &  $1.51$ & $0.86$ & $0.51$ \\ 
$3.5$ &  $2/3$ & $2.91$ &  $1.56$ & $0.83$ & $0.56$ \\ 
      &  $1$   & $3.05$ &  $1.60$ & $0.80$ & $0.59$ \\ 
\hline
\hline
\end{tabular}
\caption{Maximum neutron star mass, central baryon number density and central speed of sound  
for different values of the contact term parameters. The results for $x=0$  correspond to the case when only nucleonic TBF 
are considered. In brakets are given the corresponding values when the presence of hyperons is neglected. Masses are given 
in $M_\odot$ whereas the central baryon density, $\rho_c$, is given in fm$^{-3}$ and the central speed of sound is given in units
of $c$.}
\label{tab:tab2}
\end{table}
\end{center}
Table \ref{tab:tab2} shows the maximum neutron star mass, central baryon number density and central speed of sound for 
different values of the contact term parameters. The range of variation of these parameters allows us to explore different 
EoS and establish a lower and an upper limit to the effect of hyperonic TBF on the maximum mass.
One can get an idea of the relative importance of these forces on the EoS (and, therefore, in some way also of  
the relative weight of the phenomenological part of our calculation with respect to the microscopic one), by evaluating the ratio between 
the last four terms of Eq.\ (\ref{eq:tbf2}), and the sum $\sum_{B_i}\sum_{\vec{k}}n_{B_i}(|\vec{k}|) U_{B_i}(\vec{k})/(2{\cal V})$ 
(see Eq.\ (\ref{eq:ev})). We find that, in average, this ratio is smaller than $ 0.2$ for $\rho < 4\rho_0 $, it ranges between $0.2$ and $0.5$ 
for densities up to $ \sim 5\rho_0 $, and it is larger than $1$ for $\rho > 6 \rho_0$, clearly showing that the relative importance of hyperonic TBF
increases for larger densities.
The results for $x=0$ correspond to the case when only nucleonic TBF are considered ({\it i.e.,} $a_{YN}=b_{YN}=0$). In brakets are given the 
corresponding values when the presence of hyperons is neglected in the EoS. Note that in this case the resulting maximum mass is relatively large, 
ranging from  $2.22 M_\odot$ for $\gamma_{NN}=2$ to $2.97 M_\odot$ for $\gamma_{NN}=3.5$. The presence of hyperons induces here a reduction of the mass to 
values in the interval $1.27-1.38 M_\odot$, well below the limit of $1.4-1.5 M_\odot$. Note that the range of variation of the maximum mass 
is about $0.11 M_\odot$ in this case, compared to a range of $\sim 0.75 M_\odot$ when hyperons are absent. This is a consequence, as already 
pointed out in Ref.\ \cite{hans06}, of a strong compensation mechanism caused by the appearance of hyperons which makes the maximum mass 
quite insensitive to the pure nucleonic part of the EoS. 
As expected, the central density decreases when increasing the effect of three-body forces (the pressure is larger
and consequently the object is less compact), but at the same time the speed of sound increases, because
the EoS is stiffer. We note that when the presence of hyperons is neglected the EoS is always supraluminical. 
This is not surprising, since our approach is a non-relativistic one, and causality is, therefore, not guaranteed. 
However, note that as soon as hyperons are present in matter, the softening of the EoS induced by their presence 
is such that in these cases the EoS remains always causal.
\begin{figure}
\begin{center}
\includegraphics[width=6.0cm]{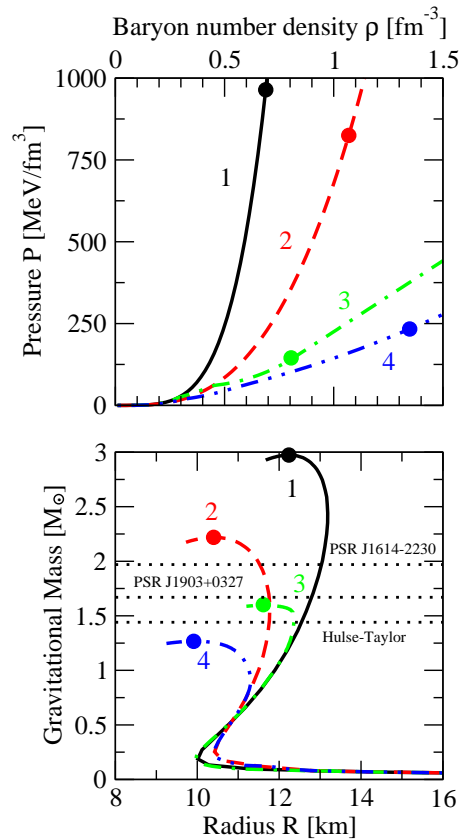}
\caption{(Color online) Upper panel: $\beta$-stable matter EoS. Lower panel: Mass-radius relation for different EoS.
Circles indicate the central baryon number density, central pressure, mass and radius of the maximum mass stellar 
configuration. Horizontal lines show the masses of the pulsars PSR J1614-2230 \cite{demorest10}, PSR J1903+0327 \cite{freire09} 
and the Hulse--Taylor one \cite{hulsetaylor}. See the text for details.}
\label{fig:fig1}
\end{center}
\end{figure}
It is clear from Table \ref{tab:tab2} that hyperonic TBF provide additional repulsion making
the EoS stiffer, and the maximum mass larger. 
For a fixed value of the exponent $\gamma_{NN}$ the maximum mass
increases when increasing $x$ ({\it i.e.,} $\gamma_{YN}$). This is an expected result, since by increasing $x$ 
we are increasing the strength of the hyperonic TBF and, as a consequence, the EoS becomes stiffer, and the 
maximum mass larger. We have checked that the rate of increase of $M_{max}$ with $x$ is slightly quadratic.
The stiffer EoS including hyperonic TBF is obtained for $\gamma_{NN}=3.5$ and $x=1$ ($\gamma_{YN}=3.05$), 
and allows for a maximum mass of about $1.60 M_\odot$. 
We note that although the inclusion of hyperonic TBF can reconcile
the maximum mass of hyperonic stars with the ``canonical'' value, they are, however, unable to make the maximum mass compatible with the observation of massive 
neutron stars, such as the recent measurements of the unusually high masses of the millisecond  pulsars PSR J1614-2230 ($1.97 \pm 0.04 M_\odot$) 
\cite{demorest10} and PSR J1903+0327 ($1.667 \pm 0.021 M_\odot$) \cite{freire09}. 
One could in principle increase arbitrarily the strength of the hyperonic TBF by 
increasing the value of the fraction $x$ in order to get a very stiff EoS which supports such large masses. However, in our opinion values of $x > 1$ give rise to 
EoS unrealistically stiff, the reason being the following: we
know that the strength of the two-body YN interaction is smaller than that of the NN one ({\it e.g.,} the single-particle potential of a $\Lambda$ in symmetric nuclear
matter at saturation for zero momentum is about $1/3$ that of the nucleons). Therefore, it is quite natural to think that probably the strength of hyperonic TBF
is either smaller or as large as the pure nucleonic ones, but not larger. Although this statement seems a bit speculative, and only a more realistic and sophisticated determination
of hyperonic TBF can give a definite answer, we consider the value of $1.60 M_\odot$ a reasonable upper limit for the maximum mass of neutron stars with hyperonic
TBF. This upper limit could be slightly larger, not more than $5 \%$, if we would have taken $a_{\Sigma N} \neq a_{\Lambda N},
b_{\Sigma N} \neq b_{\Lambda N}$, and $\gamma_{\Sigma N} \neq \gamma_{\Lambda N}$, and assumed a slightly repulsive single-particle
potential for the $\Sigma^-$ in nuclear matter at saturation density. We point out, however, that  the uncertainties related with the
binding-energy  of the $\Sigma^-$ in nuclear matter at saturation density are still very large (see {\it e.g.,} Refs. \cite{mares95, dabro99,nou02,saha04}).
Clearly, $1.27 M_\odot$ is the lower limit corresponding to the softer EoS.

Fig.\ \ref{fig:fig1} shows as a summary the EoS (upper panel) for the stiffer and the softer pure nucleonic (curves 1 and 2) and hyperonic (curves 3 and 4) 
stars, and their corresponding mass-radius relation (lower panel). Curves 1 and 2 show respectively the results for $\gamma_{NN}=3.5$ and $\gamma_{NN}=2$ 
without hyperons, while curves 3 and 4 correspond to the cases $\gamma_{NN}=3.5, x=1$, and $\gamma_{NN}=2, x=0$. Circles indicate the
value of the central baryon number density, central pressure, mass and radius of the maximum mass stellar configuration. 
The horizontal lines show the masses of the pulsars PSR J1614-2230, PSR J1903+0327 and the Hulse--Taylor one ($1.4414 \pm 0.0002$) \cite{hulsetaylor}. 
The strong softening of the EoS due to the presence 
of hyperons and the consequent reduction of the maximum mass is clearly seen. The maximum masses of hyperonic stars lay in a narrow range from $1.27$ 
to $1.60 M_\odot$, that is still compatible with the mass of Hulse-Taylor pulsar, but is well below the masses of PSR J1903+0327
and PSR J1614-2230. 

Summarizing, we use a model based on a microscopic BHF approach of hyperonic matter supplemented with additional simple phenomenological density-dependent 
contact terms to establish numerical lower and upper limits to the effect of hyperonic TBF on the maximum mass of neutron stars. 
Assuming that the strength
of these forces is either smaller or as large as the pure nucleonic ones, our results show that maximum masses of hyperonic stars lay in a narrow range from 
$1.27$ to $1.60 M_\odot$ which is still compatible with the ``canonical'' value of $1.4-1.5 M_\odot$, but it is incompatible with the observation of massive neutron
stars, such as the recent measurements of a mass of  $1.97 \pm 0.04 M_\odot$ for the millisecond pulsar PSR J1614-2230, and a mass of $1.667 \pm 0.021 M_\odot$ for 
the PSR J1903+0327 one. 
We hope that this exploratory work can serve as a motivation to perform more realistic and sophisticated 
studies of hyperonic TBF and their effects on the neutron star structure, since they have the last word on this issue.

This work has been partially supported by FCT (Portugal) under grants SFRH/BD/62353/2009 and FCOMP-01-0124-FEDER-008393 
with FCT reference CERN/FP/109316/2009, the Consolider Ingenio 2010 Programme CPAN CSD2007-00042 and Grant No. FIS2008-01661 
from MEC and FEDER (Spain) and Grant 2009GR-1289 from Generalitat de Catalunya (Spain), and by COMPSTAR, an ESF Research Networking Programme.



\end{document}